\documentclass[sigconf]{acmart}

\acmConference{}{}{}
\acmDOI{}
\acmISBN{}

\usepackage{amsmath}
\usepackage{graphicx}
\usepackage{multirow}
\usepackage{array}
\usepackage{listings}
\usepackage{xcolor}
\usepackage{float}

\definecolor{jsonkey}{RGB}{0,90,160}
\definecolor{jsonstr}{RGB}{160,60,0}

\lstdefinelanguage{json}{
  basicstyle=\ttfamily\scriptsize,
  showstringspaces=false,
  breaklines=true,
  literate=
     *{:}{{{\color{black}:}}}{1}
      {,}{{{\color{black},}}}{1}
      {\{}{{{\color{black}\{}}}{1}
      {\}}{{{\color{black}\}}}}{1}
      {[}{{{\color{black}[}}}{1}
      {]}{{{\color{black}]}}}{1},
}

\lstdefinestyle{json}{
  language=json,
  basicstyle=\ttfamily\scriptsize,
  showstringspaces=false,
  breaklines=true,
  columns=fullflexible,
}

\lstdefinestyle{spice}{
  basicstyle=\ttfamily\scriptsize,
  breaklines=true,
  showstringspaces=false,
  columns=fullflexible,
}

\begin{document}

\title{NetlistBench: Evaluating LLM Reliability in SPICE Netlist Recognition and Manipulation}

\author{Jiarui Ma}
\affiliation{%
  \institution{Southern University of Science and Technology}
  \department{School of Microelectronics}
  \city{Shenzhen}
  \country{China}
}
\email{12312626@mail.sustech.edu.cn}

\author{Jianghan Wang}
\affiliation{%
  \institution{Southern University of Science and Technology}
  \department{School of Microelectronics}
  \city{Shenzhen}
  \country{China}
}
\email{12311107@mail.sustech.edu.cn}

\author{Yuheng Ma}
\affiliation{%
  \institution{Southern University of Science and Technology}
  \department{School of Microelectronics}
  \city{Shenzhen}
  \country{China}
}
\email{12412108@mail.sustech.edu.cn}

\author{Ziyi Zhuang}
\affiliation{%
  \institution{Southern University of Science and Technology}
  \department{School of Microelectronics}
  \city{Shenzhen}
  \country{China}
}
\email{12412728@mail.sustech.edu.cn	}

\author{Xiaoguang Liu}
\affiliation{%
  \institution{Southern University of Science and Technology}
  \department{School of Microelectronics}
  \city{Shenzhen}
  \country{China}
}
\email{liuxg@sustech.edu.cn}

\renewcommand{\shortauthors}{Ma et al.}

\begin{abstract}
Large Language Models (LLMs) are increasingly used in circuit design workflows, yet their reliability on simulator-facing SPICE netlist recognition and manipulation remains poorly understood and is rarely separated from high-level design reasoning. Although netlists are textual, they encode structured circuit objects through topology and parameters. We present \textbf{NetlistBench}, a structure-verified benchmark for SPICE netlist recognition and manipulation. NetlistBench contains 2,342 cases across 24 task families, covering parameter and connectivity recognition and edits, hierarchical operations, equivalence judgment, and long-horizon compound editing. Model outputs are evaluated by a deterministic structure-aware oracle. Across six non-thinking LLMs, performance varies substantially with operation-level structural complexity. Simple local edits reach $96\%$--$100\%$ accuracy, while device addition drops to $41\%$--$83\%$ and equivalence judgment to $49\%$--$90\%$. Enabling reasoning substantially improves weaker models but does not eliminate structure-preservation failures, with performance still degrading sharply as the edit horizon increases. NetlistBench identifies netlist reliability as a distinct bottleneck for trustworthy LLM-based circuit design automation.
\end{abstract}




\maketitle

\section{Introduction}
Large language models (LLMs) are increasingly explored across the lifecycle of integrated circuit (IC) design, including domain-adapted chip-design assistance, analog circuit generation, simulation-driven optimization, and multimodal netlist extraction~\cite{pan2025surveyllmeda,chipnemo,analogcoder,toposizing,spiceassistant,netlistify,image2net}. 
By treating hardware artifacts as structured code, LLM-based systems hold the potential to accelerate electronic design automation (EDA) by translating specifications, editing topologies, and driving simulator- or layout-facing tools.

\begin{figure}[!t]
    \centering
    \includegraphics[width=\linewidth]{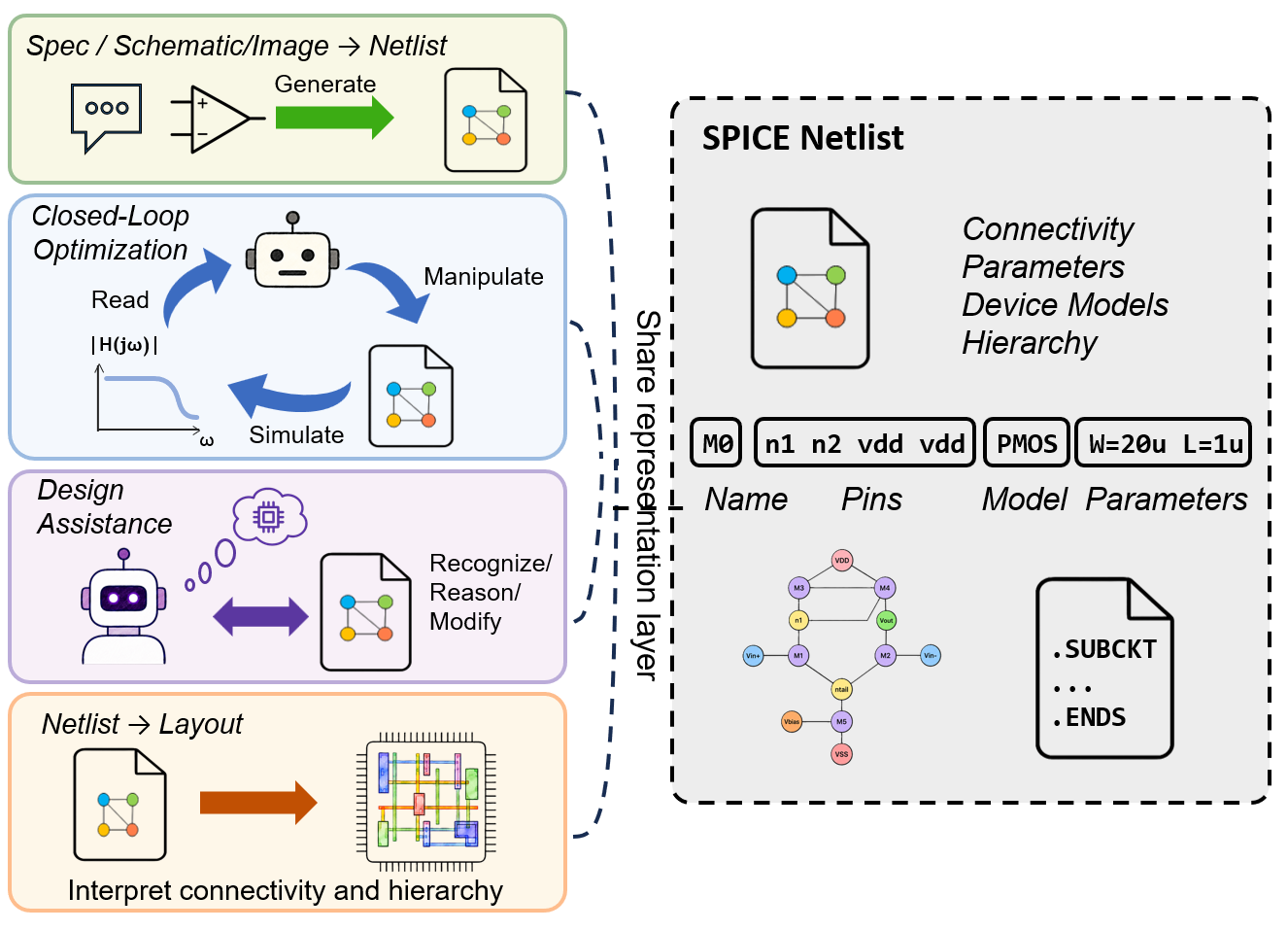}
    \caption{SPICE netlists as a common representation layer in LLM-based circuit design workflows.}
    \Description{Diagram showing SPICE netlists as an intermediate representation connecting LLM-based circuit generation, editing, optimization, simulation, schematic processing, and physical-design workflows.}
    \label{fig:gene_pip}
\end{figure}

Despite differences in their inputs, objectives, and tool interfaces, many of these workflows share SPICE netlists as a recurring representation layer (Figure~\ref{fig:gene_pip}). 
In generation-oriented settings, LLMs may synthesize netlists from specifications, schematics, or circuit images~\cite{analogcoder,netlistify,image2net,masalachai}. 
In simulation-driven optimization loops, they often revise existing netlists according to performance or simulator feedback~\cite{spiceassistant,spicepilot,toposizing}. 
In netlist-to-schematic or netlist-to-layout workflows, they may interpret connectivity, hierarchy, and device relationships to guide downstream visualization or physical design~\cite{schemato,align,gana,gnnannotation,graeb2023hierarchy}. 
Across these settings, key model actions frequently take the form of reading, generating, editing, or reasoning over netlist text.

Errors at this representation layer can directly corrupt downstream simulation, optimization, or layout reasoning. 
Consequently, failures in LLM-based circuit workflows may originate either from high-level design reasoning or from low-level netlist corruption, yet existing evaluations rarely separate these two sources.
Reliable netlist operation is therefore a prerequisite for trustworthy LLM-based circuit design workflows.

However, this prerequisite has not been directly quantified. 
General LLM-for-code benchmarks commonly evaluate executable functional
correctness through unit tests or repository test suites
~\cite{chen2021codex,jimenez2024swebench}, but do not capture
the device-specific terminal semantics, shared-node connectivity, and ordered
subcircuit interfaces of SPICE netlists.
Existing evaluations of LLM-based circuit design typically focus on end-to-end outcomes, such as syntactic validity, simulation success, specification improvement, or downstream task completion~\cite{analogcoder,spicepilot,spiceassistant}. 
Recent circuit-oriented benchmarks mainly assess domain-level capabilities, including circuit interpretation, topology reasoning, schematic understanding, AMS-domain multimodal reasoning, or graph-structured reasoning~\cite{circuit,amsbench,nlgraph}. 
While these evaluations reveal important limitations, they do not isolate the elementary operations required to interpret and modify SPICE netlists correctly. 
\textbf{Accordingly, the reliability of LLMs in performing core SPICE netlist operations remains unclear.}

To address this gap, we introduce \textbf{NetlistBench}, a structure-verified benchmark for evaluating whether LLMs can reliably recognize and manipulate analog SPICE netlists as structured circuit representations. 
NetlistBench focuses on structure-level operations that test a model's ability to recover circuit structure or apply explicit modifications without introducing unintended changes.
Model outputs are evaluated through a deterministic canonical circuit
representation that captures devices, ordered terminal bindings, node
identities, parameters, directives, and hierarchy.

The main contributions of this work are:
\begin{itemize}
    \item We formulate SPICE netlist reliability as a representation-level evaluation problem, focusing on whether LLMs can correctly recognize and manipulate netlists as structured circuit artifacts.
    
    \item We introduce \textbf{NetlistBench}, a structure-verified benchmark covering netlist structural-property recognition and natural-language-guided netlist manipulation. Using NetlistBench, we evaluate representative frontier, flash-class, and open-weight LLMs and show that reliability varies sharply across operation type and task horizon.

    \item We develop a structure-aware evaluation pipeline based on canonical circuit representations, enabling manipulation outputs to be verified beyond raw text matching or final simulation outcomes.
\end{itemize}

\section{Background}
\subsection{SPICE Netlists as Structured Circuit Representations}
\label{subsec:spice}
SPICE netlists are simulator-facing circuit descriptions that encode devices, terminals, nodes, parameters, models, ports, and subcircuit hierarchies in a compact, positional textual format~\cite{nagel1973spice,nagel1975spice2}. 
Although a netlist appears as a sequence of text lines, its underlying semantics correspond to a structured circuit object.
Each device statement typically begins with an instance prefix, followed by an ordered sequence of node names connected to specific device terminals, a model reference, and optional parameter assignments. 
For hierarchical circuits, subcircuit definitions (\texttt{.subckt}) establish ordered port interfaces, and each subcircuit instance binds external nodes to internal ports strictly according to their positional order in the instance statement.

A defining characteristic of this representation is that electrical connectivity is encoded implicitly through node-name sharing rather than explicit terminal-to-terminal links. 
For circuit simulation, node labels are sufficient because devices contribute
equations to the modified nodal analysis (MNA) system according to the nodes
attached to their terminals~\cite{ho1975mna}. Terminals sharing the same node
name are treated as electrically connected. 
For structural analysis and
manipulation, however, these connections are not represented as explicit
terminal-to-terminal links; the circuit topology must be reconstructed from
terminal--node bindings across the entire netlist.

These properties make netlist operations different from ordinary text editing. A correct edit must preserve terminal-role bindings, maintain consistent node identities, and avoid unintended changes to unrelated devices or subcircuit interfaces. NetlistBench therefore evaluates netlist outputs through structural equivalence to a canonical circuit representation rather than through surface-string similarity.

\subsection{Reported Limitations of LLMs on SPICE Netlists}

Existing studies have reported several limitations of LLMs in processing circuit representations and SPICE-like netlists. 
At the circuit-reasoning level, benchmarks such as CIRCUIT and AMSbench show that LLMs can struggle with topology-heavy circuit interpretation and multi-step circuit reasoning~\cite{circuit,amsbench}. 
At the generation and adaptation level, systems such as SPICEPilot, SPICEAssistant, AnalogCoder, and Spice Wizard rely on simulation feedback, syntax checks, or tool-assisted repair loops to improve SPICE code or netlist generation, indicating that unvalidated LLM outputs may not provide sufficient reliability for direct downstream use~\cite{spicepilot,spiceassistant,analogcoder,spicewizard}. 
At the netlist-analysis level, SPICED studies LLM-aided detection and localization of syntactical bugs and analog Trojans in A/MS netlists, further showing that node, parameter, subcircuit, and connectivity errors are meaningful failure classes in SPICE-like representations~\cite{spiced}.

Complementary representation-oriented work suggests why these failures are difficult to avoid with ordinary text modeling alone. 
CircuitFormer highlights the mismatch between standard language tokenization and the graph-structured semantics of circuits, while Image2Net evaluates diagram-to-netlist conversion using graph-structured netlist comparison rather than raw string matching~\cite{circuitformer,image2net}. 
Taken together, these findings suggest that LLM failures on SPICE netlists are not merely surface-level syntax issues, but are closely related to the difficulty of preserving structured circuit semantics in a positional textual representation.

However, these studies evaluate netlists within broader generation,
simulation, conversion, or reasoning pipelines rather than isolating
representation-level netlist operations.

\section{Benchmark Design}

NetlistBench evaluates representation-level netlist reliability through two modalities: recognition, which extracts or compares circuit structure, and manipulation, which applies explicit natural-language edits to SPICE netlists. This design separates structure interpretation and structure-preserving transformation from high-level design reasoning, simulator behavior, and optimization. 
\begin{table}[H]
\centering
\caption{Composition of the NetlistBench source corpus.}
\label{tab:source-corpus}

\small
\setlength{\tabcolsep}{3.5pt}
\renewcommand{\arraystretch}{1.08}

\begin{tabular}{
    @{}
    l
    r
    >{\raggedright\arraybackslash}p{3.05cm}
    @{}
}
\toprule
Source subset & Count & Structural summary \\
\midrule

AnalogGenie--Simple
& 492
& Median: 5 devices \\

AnalogGenie--Medium
& 1,752
& Median: 20 devices \\

AnalogGenie--Complex
& 594
& Median: 36 devices; maximum: 69 \\

\cmidrule(lr){1-3}
\textit{Flat subtotal}
& \textbf{2,838}
& Approximately 58,000 device instances \\

\addlinespace[2pt]
ALIGN hierarchical
& 931
& 21 topology families; 3--4 subcircuits per netlist \\

\midrule
\textbf{Total}
& \textbf{3,769}
& \\

\bottomrule
\end{tabular}
\end{table}

\subsection{Source Corpus}

NetlistBench uses two complementary SPICE netlist sources, summarized in
Table~\ref{tab:source-corpus}. AnalogGenie provides flat CMOS analog netlists
originally developed for topology discovery~\cite{analoggenie}; after
normalization, these circuits are used to construct flat recognition and
manipulation tasks. ALIGN provides hierarchical analog netlists from a layout
automation flow~\cite{align}; its circuits contain multiple
\texttt{.subckt} definitions and top-level instance calls and are used for
hierarchical tasks.

\subsection{Instance Construction Pipeline}

NetlistBench constructs evaluation instances through a deterministic, template-driven pipeline rather than stochastic or unconstrained generation. 
Starting from source SPICE netlists, the pipeline applies family-specific transformation rules, syntax perturbations, and prompt templates to produce reproducible benchmark cases. 
Each instance is represented as a self-contained triplet:
\begin{equation}
\mathcal{I}_i =
\left(
\mathbf{N}^{(i)}_{\mathrm{src}},
\mathbf{T}^{(i)}_{\mathrm{inst}},
\mathbf{Y}^{(i)}_{\mathrm{target}}
\right),
\end{equation}
where $\mathbf{N}^{(i)}_{\mathrm{src}}$ denotes the source SPICE netlist, 
$\mathbf{T}^{(i)}_{\mathrm{inst}}$ denotes the explicit task instruction, and 
$\mathbf{Y}^{(i)}_{\mathrm{target}}$ denotes the task-specific target used for evaluation. 
Both $\mathbf{T}^{(i)}_{\mathrm{inst}}$ and $\mathbf{Y}^{(i)}_{\mathrm{target}}$ are produced by deterministic, family-specific templates, ensuring that each instance has an unambiguous instruction and reproducible ground truth.

The form of $\mathbf{Y}^{(i)}_{\mathrm{target}}$ depends on the task type. 
For manipulation tasks, it is the uniquely determined target SPICE netlist that realizes the requested structural transformation. 
For recognition tasks, it is the canonical JSON answer derived from the source netlist. 
For equivalence judgment tasks, it is the binary structural-equivalence verdict.

All instances undergo automated construction-time validation before inclusion.
The checker ensures that manipulation targets implement exactly the specified
structural changes without unintended edits, and that recognition and
equivalence targets are consistent with the canonical IR of the corresponding
input netlist or netlist pair.

\begin{figure}[t]
    \centering
    \includegraphics[width=\columnwidth]{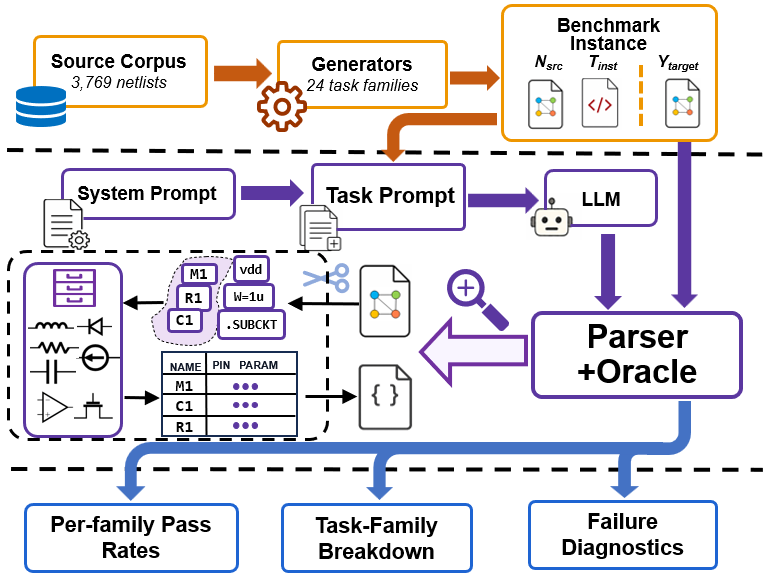}
     \caption{NetlistBench pipeline for generating benchmark cases and evaluating model outputs with a structure-aware oracle.}
    \Description{Pipeline diagram showing source netlists, deterministic benchmark-instance construction, model inference, canonical circuit representation, and structure-aware evaluation.}
    \label{fig:pipeline}
\end{figure}

\begin{figure*}[t]
    \centering
    \includegraphics[width=\linewidth]{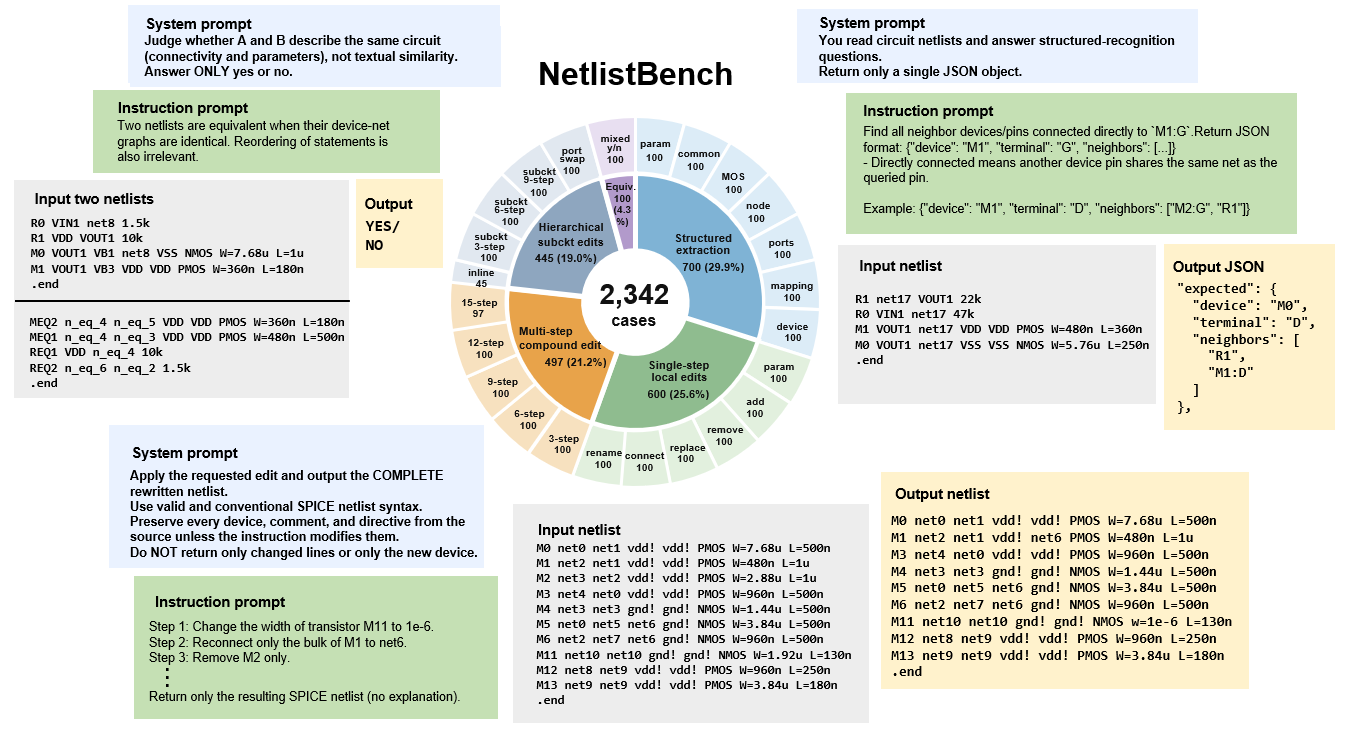}
    \caption{Overview of the NetlistBench benchmark, showing the distribution of cases across task families and representative task examples.}
    \Description{Overview of the recognition and manipulation task families in NetlistBench, including their case distribution and representative netlist tasks.}
    \label{fig:family}
\end{figure*}
\subsection{Structure-Aware Evaluation Oracle}
\label{sec:oracle}
This structure-aware evaluation follows the broader principle that circuit artifacts should be compared through their underlying connectivity and device structure rather than by surface text. 
Similar graph-structured evaluation ideas have been used in diagram-to-netlist conversion, where generated and reference netlists are compared through heterogeneous circuit graphs rather than raw strings~\cite{image2net}.

Concretely, each model output and the reference target are parsed into the canonical IR 
---a normalized structure that lists every device by instance name with its device kind, ordered terminal nodes, and parameters, together with top-level directives and, for hierarchical circuits, each subcircuit's port interface and internal devices. 
An output passes only if its IR matches the reference IR under a fixed set of semantics-preserving normalizations: the two must contain the same set of named devices, with no missing or extra device, identical terminal-node bindings, parameters equal up to numeric normalization (e.g., \texttt{1k} equals \texttt{1000}), and identical top-level directives; symmetric two-terminal passives ($R$/$C$/$L$) are compared with unordered terminals, and subcircuit definitions must agree on their port interface (port order treated as semantic except for extraction tasks), internal devices, and directives. 
This exact-match-up-to-normalization rule directly encodes the name-preservation and locality constraints of edit tasks: renaming an untouched node, dropping or duplicating a device, or perturbing an unrelated parameter each surfaces as an IR mismatch and fails the case. 
The same canonical-IR comparator scores SPICE and PySpice~\cite{pyspice} outputs through a uniform interface.

For the equivalence-judgment family, the task is instead to decide whether two netlists denote the same circuit up to consistent node and instance renaming. 
To validate the generated ground-truth labels for this family, we projected the IR into a labeled bipartite device--net graph---device nodes labeled by type, model, and normalized parameter signatures, and device--net edges labeled by terminal roles---and ran a VF2 graph-isomorphism check~\cite{cordella2004vf2} confirming that positive pairs are isomorphic and negative pairs are not. 
This isomorphism check audits equivalence-case labels only and is not part of scoring model outputs: scoring must instead preserve device and node names, whereas VF2 equates circuits \emph{up to} renaming and would therefore mask the very name- and locality-violations that the edit tasks are designed to detect.

\subsection{Task Families}

NetlistBench contains 24 task families across recognition and manipulation, as
summarized in Figure~\ref{fig:family}. The recognition modality contains 800
cases across eight families. Seven evaluate structured extraction of device
parameters, terminal connectivity, node incidence, subcircuit interfaces, and
instance mappings, while the eighth evaluates structural equivalence between
netlist pairs.

The manipulation modality contains 1,542 cases across 16 families. Six
single-edit families cover connectivity editing, device addition, removal and
replacement, parameter editing, and rename propagation. Five flat compound
families combine 3, 6, 9, 12, or 15 dependent operations, and five hierarchical
families evaluate subcircuit expansion, interface modification, and multi-step
internal editing.

\section{Evaluation}

\subsection{Experimental Setup and Protocol}

We evaluate six single-shot non-thinking models spanning frontier, flash-class, and open-weight tiers: Claude Sonnet 4.6, GPT-4.1, Gemini 2.5 Flash, DeepSeek-V4-Flash, Qwen3.6-Flash, and Qwen3-30B-A3B. 
All are queried through official provider APIs with explicit reasoning modes disabled, so the main comparison measures base netlist-operation reliability rather than reasoning elicitation. 
As a reasoning reference, we additionally evaluate the same DeepSeek-V4-Flash with its native thinking mode enabled; this column is reported separately and excluded from the non-thinking comparison. 
We also run two controlled secondary analyses on a paired stratified subset: SPICE versus PySpice output representation, and direct prompting versus native thinking and CoT prompting~\cite{wei2022cot}.

Each model is queried once per case with deterministic decoding, and retries are used only for transport failures. 
Responses are graded by the structure-aware oracle in Section~\ref{sec:oracle} and reduced to binary pass/fail outcomes: manipulation outputs must match the reference structure, recognition outputs must match the canonical JSON answer, and equivalence judgments must match the reference verdict. 
The parser tolerates incidental code fences, but empty, unparsable, or structurally invalid outputs fail. 
We report pass rates with Wilson 95\% confidence intervals~\cite{wilson1927probable}; Table~\ref{tab:leaderboard} gives the per-family case count $n$, and aggregate ablation intervals are stated explicitly. 
Subtotals and overall scores are case-weighted. 
Reasoning-mode results are single samples and may carry run-to-run variance.

\paragraph{Availability.}
NetlistBench is publicly available at \url{https://github.com/WoshiMayou/NetlistBench}. 
The repository contains all 2,342 benchmark cases across the 24 task families, the deterministic structure-aware oracle, seeded case-generation scripts, evaluation runners, and the per-family prompt templates required to reproduce the benchmark evaluation. 
The code is released under the Apache-2.0 license, while the benchmark cases and prompts are released under CC BY 4.0.

\begin{table}[t]
\centering
\small
\setlength{\tabcolsep}{3.4pt}
\renewcommand{\arraystretch}{0.96}
\caption{Per-family NetlistBench pass rates (\%). C/G/Ge/QF/Q30/DS denote Claude-S4.6, GPT-4.1, Gemini-2.5-F, Qwen3.6-F, Qwen3-30B, and DeepSeek-V4-F. DS+R$^\dagger$ is the reasoning reference, excluded from bolding; underlines exceed all non-thinking models.}
\label{tab:leaderboard}
\begin{tabular}{@{}l r cccccc|c@{}}
\toprule
Task & $n$ & C & G & Ge & QF & Q30 & DS & DS+R$^\dagger$ \\
\midrule

Conn. edit & 100 & \textbf{97} & 91 & 93 & 59 & 47 & 85 & \underline{99} \\
Dev. add & 100 & \textbf{83} & 72 & 57 & 54 & 43 & 41 & 55 \\
Dev. remove & 100 & \textbf{100} & \textbf{100} & \textbf{100} & 98 & 97 & \textbf{100} & 100 \\
Dev. replace & 100 & \textbf{95} & 93 & 86 & 75 & 61 & 78 & 91 \\
Param. edit & 100 & \textbf{100} & 99 & 98 & 99 & 96 & 98 & 99 \\
Rename prop. & 100 & \textbf{99} & 98 & 97 & 85 & 89 & 92 & \underline{100} \\
Comp. 3 & 100 & \textbf{80} & 71 & 70 & 34 & 28 & 44 & 74 \\
Comp. 6 & 100 & 57 & \textbf{58} & 33 & 6 & 1 & 18 & \underline{63} \\
Comp. 9 & 100 & \textbf{56} & 51 & 21 & 2 & 1 & 6 & 50 \\
Comp. 12 & 100 & \textbf{41} & 39 & 17 & 0 & 0 & 1 & 33 \\
Comp. 15 & 97 & 26 & \textbf{34} & 6 & 0 & 0 & 0 & 31 \\
Subckt inline & 45 & 56 & \textbf{62} & 33 & 16 & 2 & 42 & \underline{98} \\
Port swap & 100 & \textbf{88} & 80 & 83 & 33 & 19 & 65 & \underline{97} \\
Subckt comp. 3 & 100 & \textbf{79} & 69 & 55 & 39 & 21 & 32 & 66 \\
Subckt comp. 6 & 100 & \textbf{67} & 55 & 36 & 5 & 4 & 8 & 54 \\
Subckt comp. 9 & 100 & 42 & \textbf{43} & 41 & 1 & 1 & 3 & \underline{47} \\
\midrule
\textit{Edit subtotal} & 1542 & \textbf{74} & 70 & 59 & 39 & 33 & 45 & 71 \\
\midrule
Dev. param. & 100 & 99 & \textbf{100} & \textbf{100} & \textbf{100} & 99 & \textbf{100} & 100 \\
Sem. term. conn. & 100 & \textbf{99} & 73 & 74 & 25 & 6 & 82 & \underline{100} \\
Ord. term. conn. & 100 & \textbf{100} & \textbf{100} & 99 & 99 & \textbf{100} & \textbf{100} & 100 \\
Node inc. & 100 & \textbf{98} & 42 & 59 & 21 & 15 & 53 & 96 \\
Subckt ports & 100 & \textbf{100} & \textbf{100} & \textbf{100} & \textbf{100} & 99 & \textbf{100} & 100 \\
Inst. port map & 100 & \textbf{100} & 96 & 87 & 89 & 65 & 96 & 100 \\
Term. neigh. inc. & 100 & \textbf{97} & 13 & 20 & 4 & 2 & 12 & 93 \\
Equiv. judge & 100 & \textbf{90} & 67 & 61 & 55 & 56 & 49 & \underline{97} \\
\midrule
\textit{Recog. subtotal} & 800 & \textbf{98} & 74 & 75 & 62 & 55 & 74 & 98 \\
\midrule
\textit{Overall} & 2342 & \textbf{82} & 71 & 64 & 47 & 41 & 55 & 81 \\
\bottomrule
\end{tabular}
\end{table}

\subsection{Performance Across Task Families}

Table~\ref{tab:leaderboard} shows substantial variation across operation
types. Local operations that primarily modify explicit text are the most
reliable: device removal and parameter editing reach $96\%$--$100\%$ across
models. Reliability decreases for operations that require maintaining
connectivity or introducing new structure, including connectivity editing,
device replacement, device addition, subcircuit port swapping, and inline
expansion.

Recognition exhibits a similar distinction. Explicit attributes such as device
parameters, ordered terminal lists, and subcircuit ports are extracted with
high accuracy, whereas relational queries vary substantially across models.
Node incidence ranges from $15\%$ to $98\%$, semantic terminal connectivity
from $6\%$ to $99\%$, and terminal-neighbor incidence from $2\%$ to $97\%$.
Structural equivalence judgment also remains challenging, with pass rates from
$49\%$ to $90\%$.

Overall non-thinking pass rates range from $41\%$ to $82\%$. The results
indicate that current models are considerably more reliable on localized
attribute extraction and substitution than on recovering or preserving the
implicit connectivity graph. Enabling reasoning raises DeepSeek-V4-Flash from
$55\%$ to $81\%$, but does not consistently surpass the strongest non-thinking
model.

\begin{figure}[t]
    \centering
    \includegraphics[
        width=\linewidth,
        keepaspectratio
    ]{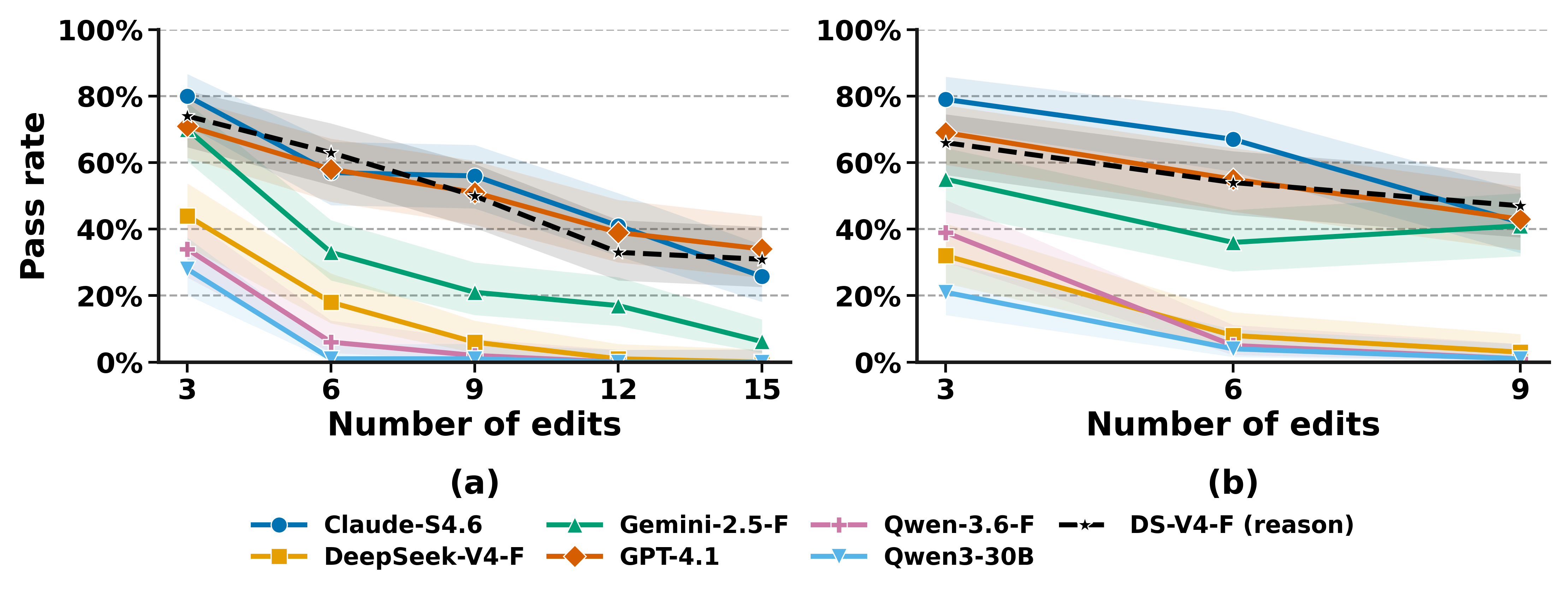}
    \caption{Pass rates under compound editing with increasing numbers of
    dependent edits for (a) flat netlists and (b) hierarchical subcircuit
    netlists. Lines show the mean pass rates, and shaded regions indicate
    the corresponding variation.}
    \Description{Two compound-editing plots showing pass rate versus the number of dependent edits for flat and hierarchical netlists. All model curves decrease as the edit horizon grows.}
    \label{fig:compound_lines}
\end{figure}
\subsection{Long-Horizon Compound Editing}
\label{sec:longhorizon}

The compound editing families chain $3$, $6$, $9$, $12$, and $15$ mutually dependent edits into a single instruction, revealing a substantial reliability degradation in NetlistBench.
Accuracy declines consistently as the edit horizon increases (Figure~\ref{fig:compound_lines}). 
Even models with strong short-horizon performance degrade substantially: Claude drops from $80\%$ at 3 steps to $26\%$ at 15 steps, while GPT-4.1 drops from $71\%$ to $34\%$. 
The remaining models decline to near-zero accuracy at longer horizons, with Gemini decreasing from $70\%$ to $6\%$, DeepSeek-V4-Flash from $44\%$ to $0\%$, and both Qwen3.6-Flash and the open-weight Qwen3-30B from $34\%$/$28\%$ to $0\%$. 
Crucially, reasoning does not eliminate this trend: DeepSeek-V4-Flash with reasoning enabled, although far stronger at short horizons ($74\%$ at 3 steps), still falls to $31\%$ at 15 steps.

The degradation is not simply a consequence of weak atomic editing. 
Long-horizon compound tasks require models to track multiple dependent edit intents, update intermediate circuit state, and preserve edit locality across an extended instruction sequence. 
Because the edits are mutually dependent, errors compound across the sequence: even a high per-edit success rate yields a low joint success probability once many edits must all be correct. 
The same downward trend appears in the hierarchical compound family, indicating that this effect is not limited to flat netlists. 
These results show that high reliability on isolated edits does not translate into reliable multi-step netlist transformation.
This pattern is consistent with broader observations that small per-step error rates can compound sharply over long execution horizons~\cite{illusiondiminishingreturns}.

\subsection{Representation and Reasoning}
\label{sec:repr-reason}
We study two mitigations on DeepSeek-V4-Flash and Qwen3.6-Flash: changing the circuit representation (SPICE$\rightarrow$PySpice) and enabling explicit reasoning (native thinking and CoT prompting~\cite{wei2022cot}). 
Both analyses use the same paired, stratified subset ($30$ cases per family, 28 for Comp.\ 15, $n{=}718$); Table~\ref{tab:ablation} reports overall pass rates with Wilson 95\% intervals, and paired arms are compared with McNemar's test.

On this subset, reasoning gives the larger aggregate gains: native thinking raises both models by roughly $30$--$40$ points ($p<10^{-6}$), while CoT also improves performance but less strongly. 
The task-level results in Table~\ref{tab:prompting_modes} and Figure~\ref{fig:radar_dtc_ablation} show similar gains across several structure-heavy families. 
By contrast, PySpice has a smaller and less consistent effect: it improves DeepSeek-V4-Flash overall ($p<0.001$), but not Qwen3.6-Flash ($p=0.51$), with task-level trends shown in Table~\ref{tab:spice_pyspice_results} and Figure~\ref{fig:compound_dtc_ablation}. 

Because each per-family cell has only $30$ cases, we treat task-level patterns as descriptive. 
Overall, reasoning is the stronger mitigation here, but errors still concentrate on long-horizon compound edits, hierarchical operations, and relational structural queries; neither mitigation makes current LLMs sufficiently reliable for unverified netlist editing.
\begin{table}[t]
\centering
\small
\setlength{\tabcolsep}{3.5pt}
\renewcommand{\arraystretch}{1.05}
\caption{Mitigation results on the paired subset ($n{=}718$), with Wilson 95\% confidence intervals.}
\label{tab:ablation}
\begin{tabular}{@{}lcccc@{}}
\toprule
Model & Direct & Native & CoT & PySpice \\
\midrule
DeepSeek-V4-F & $52$ & $81$ & $75$ & $58$ \\
 & $[48,55]$ & $[78,83]$ & $[72,78]$ & $[55,62]$ \\
Qwen3.6-F & $45$ & $85$ & $78$ & $45$ \\
 & $[41,48]$ & $[82,88]$ & $[75,81]$ & $[42,49]$ \\
\bottomrule
\end{tabular}
\end{table}

\section{Discussion}
\subsection{Implications for LLM-Based Netlist Editing}
NetlistBench shows that netlist reliability cannot be reduced to general circuit knowledge or output-format compliance. 
Models tend to perform better on localized edits, such as parameter changes and device removal, while showing reduced reliability on tasks involving structural attachment, ordered port handling, equivalence judgment, or multi-step edits.
The observed failures are frequently structural rather than purely procedural: recognition outputs usually follow the required JSON schema but contain incorrect circuit facts, while manipulation failures involve omitted edits, duplicated edits, loss of locality, unintended terminal rebinding, and topology drift.

This error pattern follows directly from the representation properties described in Section~\ref{subsec:spice}. Local substitutions and deletions often require only limited changes to already explicit text, while attachment, hierarchy, equivalence, and compound editing require the model to maintain an implicit circuit graph across terminal roles, node identities, and subcircuit interfaces. The observed failures therefore indicate a structure-preservation bottleneck: models can often produce syntactically plausible netlists, but still lose edit locality, perturb unrelated bindings, or fail to maintain consistent topology across multiple dependent operations.

These results suggest that current LLMs should not be treated as standalone, unverified netlist editors. 
Reasoning modes and alternative surface representations can mitigate some failures, but neither fully resolves the structure-preservation problem. 
More robust workflows may need to decompose complex edits, verify each intermediate netlist structurally, and provide feedback when unintended changes are detected. 
Future work may also explore graph-based or other structured circuit representations that expose connectivity more directly than raw SPICE text.

\subsection{Limitations}
NetlistBench evaluates bounded, circuit-block-level netlists rather than
industrial-scale post-layout decks. Although the compound tasks increase the
number of dependent edits, all evaluated netlists fit within the tested
models' context windows. The benchmark therefore does not assess long-context
retrieval, hierarchical partitioning, or direct processing of extracted
netlists containing millions of device and parasitic statements.

The current release also covers a restricted circuit and syntax domain,
primarily flat and hierarchical analog CMOS blocks from AnalogGenie and ALIGN.
It does not comprehensively cover symbolic \texttt{.param} expressions,
complex \texttt{.model} cards, behavioral or controlled sources, include
hierarchies, extracted parasitics, or simulator- and PDK-specific syntax.
The reported results therefore should not be assumed to transfer unchanged
to RF, power, digital, or mixed-signal netlists.

Finally, task instructions are generated from deterministic templates to
isolate structural capabilities and enable unambiguous grading. They do not
capture the full linguistic variability or design intent of real
designer--assistant interactions. Each model--case pair is evaluated once,
so the results characterize the tested API snapshots rather than complete
output distributions.

\section{Conclusion}

NetlistBench shows that netlist reliability is a distinct bottleneck for LLM-based circuit design. 
Current models often handle local substitutions and simple extraction, but remain fragile on connectivity-sensitive edits, hierarchy manipulation, structural equivalence, and long-horizon compound transformations. 
Reasoning improves performance, yet does not make LLMs reliable unverified editors of simulator-facing netlists. 
These findings motivate decomposed editing workflows, structure-aware verification after each edit, and circuit representations that expose topology more directly than raw SPICE text.

\clearpage
\bibliographystyle{ACM-Reference-Format}
\bibliography{references}

@article{pan2025surveyllmeda,
  author    = {Pan, Jingyu and Zhou, Guanglei and Chang, Chen-Chia and Jacobson, Isaac and Hu, Jiang and Chen, Yiran},
  title     = {A Survey of Research in Large Language Models for Electronic Design Automation},
  journal   = {ACM Transactions on Design Automation of Electronic Systems},
  year      = {2025},
  volume    = {30},
  number    = {3},
  articleno = {34},
  numpages  = {21},
  doi       = {10.1145/3715324},
  publisher = {Association for Computing Machinery}
}

@misc{chipnemo,
  author        = {Liu, Mingjie and Ene, Teodor-Dumitru and Kirby, Robert and Cheng, Chris and Pinckney, Nathaniel and Liang, Rongjian and Alben, Jonah and Anand, Himyanshu and Banerjee, Sanmitra and Bayraktaroglu, Ismet and others},
  title         = {{ChipNeMo}: Domain-Adapted {LLM}s for Chip Design},
  year          = {2023},
  eprint        = {2311.00176},
  archivePrefix = {arXiv},
  primaryClass  = {cs.CL}
}

@inproceedings{analogcoder,
  author    = {Lai, Yao and Lee, Sungyoung and Chen, Guojin and Poddar, Souradip and Hu, Mengkang and Pan, David Z. and Luo, Ping},
  title     = {{AnalogCoder}: Analog Circuit Design via Training-Free Code Generation},
  booktitle = {Proceedings of the AAAI Conference on Artificial Intelligence},
  year      = {2025},
  volume    = {39},
  number    = {1},
  pages     = {379--387},
  doi       = {10.1609/aaai.v39i1.32016}
}

@misc{toposizing,
  author        = {Wei, Ziming and Kong, Zichen and Wang, Yuan and Pan, David Z. and Tang, Xiyuan},
  title         = {{TopoSizing}: An {LLM}-aided Framework of Topology-based Understanding and Sizing for {AMS} Circuits},
  year          = {2025},
  eprint        = {2509.14169},
  archivePrefix = {arXiv},
  primaryClass  = {cs.LG}
}

@misc{spiceassistant,
  author        = {Nau, Simon and Krummenauer, Jan and Zimmermann, Andr{\'e}},
  title         = {Evaluating {LLM}-based Workflows for Switched-Mode Power Supply Design},
  year          = {2025},
  eprint        = {2507.10639},
  archivePrefix = {arXiv},
  primaryClass  = {cs.AR}
}

@inproceedings{netlistify,
  author    = {Huang, Chun-Yen and Chen, Hsuan-I and Ho, Hao-Wen and Kang, Pei-Hsin and Lin, Mark Po-Hung and Liu, Wen-Hao and Ren, Haoxing},
  title     = {{Netlistify}: Transforming Circuit Schematics into Netlists with Deep Learning},
  booktitle = {Proceedings of the 2025 ACM/IEEE 7th Symposium on Machine Learning for CAD (MLCAD)},
  year      = {2025},
  pages     = {1--8},
  doi       = {10.1109/MLCAD65511.2025.11189145},
  publisher = {IEEE}
}

@misc{image2net,
  author        = {Xu, Haohang and Liu, Chengjie and Wang, Qihang and Huang, Wenhao and Xu, Yongjian and Chen, Weiyu and Peng, Anlan and Li, Zhijun and Li, Bo and Qi, Lei and Yang, Jun and Du, Yuan and Du, Li},
  title         = {{Image2Net}: Datasets, Benchmark and Hybrid Framework to Convert Analog Circuit Diagrams into Netlists},
  year          = {2025},
  eprint        = {2508.13157},
  archivePrefix = {arXiv},
  primaryClass  = {cs.AR}
}

@misc{masalachai,
  author        = {Bhandari, Jitendra and Bhat, Vineet and He, Yuheng and Garg, Siddharth and Rahmani, Hamed and Karri, Ramesh},
  title         = {{Masala-CHAI}: A Large-Scale {SPICE} Netlist Dataset for Analog Circuits by Harnessing {AI}},
  year          = {2024},
  eprint        = {2411.14299},
  archivePrefix = {arXiv},
  primaryClass  = {cs.AR}
}

@misc{schemato,
  author        = {Matsuo, Ryoga and Uhlich, Stefan and Venkitaraman, Arun and Bonetti, Andrea and Hsieh, Chia-Yu and Momeni, Ali and Mauch, Lukas and Capone, Augusto and Ohbuchi, Eisaku and Servadei, Lorenzo},
  title         = {{Schemato}: An {LLM} for Netlist-to-Schematic Conversion},
  year          = {2024},
  eprint        = {2411.13899},
  archivePrefix = {arXiv},
  primaryClass  = {cs.LG}
}

@misc{spicepilot,
  author        = {Vungarala, Deepak and Alam, Sakila and Ghosh, Arnob and Angizi, Shaahin},
  title         = {{SPICEPilot}: Navigating {SPICE} Code Generation and Simulation with {AI} Guidance},
  year          = {2024},
  eprint        = {2410.20553},
  archivePrefix = {arXiv},
  primaryClass  = {cs.AR}
}

@article{align,
  author    = {Dhar, Tonmoy and Kunal, Kishor and Li, Yaguang and Madhusudan, Meghna and Poojary, Jitesh and Sharma, Arvind K. and Xu, Wenbin and Burns, Steven M. and Harjani, Ramesh and Hu, Jiang and Kirkpatrick, Desmond A. and Mukherjee, Parijat and Yaldiz, Soner and Sapatnekar, Sachin S.},
  title     = {{ALIGN}: A System for Automating Analog Layout},
  journal   = {IEEE Design \& Test},
  year      = {2021},
  volume    = {38},
  number    = {2},
  pages     = {8--18},
  doi       = {10.1109/MDAT.2020.3042177}
}

@inproceedings{gana,
  author    = {Kunal, Kishor and Dhar, Tonmoy and Madhusudan, Meghna and Poojary, Jitesh and Sharma, Arvind K. and Xu, Wenbin and Burns, Steven M. and Hu, Jiang and Harjani, Ramesh and Sapatnekar, Sachin S.},
  title     = {{GANA}: Graph Convolutional Network Based Automated Netlist Annotation for Analog Circuits},
  booktitle = {Proceedings of the 2020 Design, Automation \& Test in Europe Conference \& Exhibition (DATE)},
  year      = {2020},
  pages     = {55--60},
  doi       = {10.23919/DATE48585.2020.9116329},
  publisher = {IEEE}
}

@article{gnnannotation,
  author    = {Kunal, Kishor and Dhar, Tonmoy and Madhusudan, Meghna and Poojary, Jitesh and Sharma, Arvind K. and Xu, Wenbin and Burns, Steven M. and Hu, Jiang and Harjani, Ramesh and Sapatnekar, Sachin S.},
  title     = {{GNN}-Based Hierarchical Annotation for Analog Circuits},
  journal   = {IEEE Transactions on Computer-Aided Design of Integrated Circuits and Systems},
  year      = {2023},
  volume    = {42},
  number    = {9},
  pages     = {2801--2814},
  doi       = {10.1109/TCAD.2023.3236269}
}

@inproceedings{graeb2023hierarchy,
  author    = {Graeb, Helmut and Leibl, Markus},
  title     = {Learning from the Implicit Functional Hierarchy in an Analog Netlist},
  booktitle = {Proceedings of the 2023 ACM International Symposium on Physical Design (ISPD)},
  year      = {2023},
  pages     = {93--100},
  doi       = {10.1145/3569052.3578921},
  publisher = {Association for Computing Machinery}
}

@techreport{nagel1973spice,
  author      = {Nagel, Laurence W. and Pederson, Donald O.},
  title       = {{SPICE} (Simulation Program with Integrated Circuit Emphasis)},
  institution = {Electronics Research Laboratory, University of California, Berkeley},
  year        = {1973},
  number      = {UCB/ERL M382},
  url         = {https://www2.eecs.berkeley.edu/Pubs/TechRpts/1973/22871.html}
}

@techreport{nagel1975spice2,
  author      = {Nagel, Laurence W.},
  title       = {{SPICE2}: A Computer Program to Simulate Semiconductor Circuits},
  institution = {Electronics Research Laboratory, University of California, Berkeley},
  year        = {1975},
  number      = {UCB/ERL M520},
  url         = {https://www2.eecs.berkeley.edu/Pubs/TechRpts/1975/9602.html}
}

@article{ho1975mna,
  author  = {Ho, Chung-Wen and Ruehli, Albert E. and Brennan, Pierce A.},
  title   = {The Modified Nodal Approach to Network Analysis},
  journal = {IEEE Transactions on Circuits and Systems},
  year    = {1975},
  volume  = {22},
  number  = {6},
  pages   = {504--509},
  doi     = {10.1109/TCS.1975.1084079}
}

@misc{circuit,
  author        = {Skelic, Lejla and Xu, Yan and Cox, Matthew and Lu, Wenjie and Yu, Tao and Han, Ruonan},
  title         = {{CIRCUIT}: A Benchmark for Circuit Interpretation and Reasoning Capabilities of {LLM}s},
  year          = {2025},
  eprint        = {2502.07980},
  archivePrefix = {arXiv},
  primaryClass  = {cs.LG}
}

@misc{amsbench,
  author        = {Shi, Yichen and Zhang, Ze and Wang, Hongyang and Tao, Zhuofu and Li, Zhongyi and Chen, Bingyu and Wang, Yaxin and Yu, Zhiping and Lin, Ting-Jung and He, Lei},
  title         = {{AMSbench}: A Comprehensive Benchmark for Evaluating {MLLM} Capabilities in {AMS} Circuits},
  year          = {2025},
  eprint        = {2505.24138},
  archivePrefix = {arXiv},
  primaryClass  = {cs.LG}
}

@inproceedings{nlgraph,
  author    = {Wang, Heng and Feng, Shangbin and He, Tianxing and Tan, Zhaoxuan and Han, Xiaochuang and Tsvetkov, Yulia},
  title     = {Can Language Models Solve Graph Problems in Natural Language?},
  booktitle = {Advances in Neural Information Processing Systems},
  year      = {2023},
  volume    = {36},
  pages     = {30840--30861}
}

@misc{spicewizard,
  author       = {Divakar, Aakash and Anekar, Aditya and Kulkarni, Manas},
  title        = {Spice Wizard: A Unified {AI} Agent for Netlist Generation},
  year         = {2026},
  publisher    = {TechRxiv},
  doi          = {10.36227/techrxiv.177162431.10627206},
  note         = {Preprint}
}

@misc{spiced,
  author        = {Chaudhuri, Jayeeta and Thapar, Dhruv and Chaudhuri, Arjun and Firouzi, Farshad and Chakrabarty, Krishnendu},
  title         = {{SPICED}: Syntactical Bug and Trojan Pattern Identification in {A/MS} Circuits using {LLM}-Enhanced Detection},
  year          = {2024},
  eprint        = {2408.16018},
  archivePrefix = {arXiv},
  primaryClass  = {cs.AR}
}

@misc{circuitformer,
  author        = {Islam, Md Touhidul and Saha, Sujan Kumar and Farahmandi, Farimah and Tehranipoor, Mark},
  title         = {{CircuitFormer}: A Circuit Language Model for Analog Topology Design from Natural Language Prompt},
  year          = {2026},
  eprint        = {2605.05773},
  archivePrefix = {arXiv},
  primaryClass  = {cs.AR}
}

@inproceedings{analoggenie,
  author    = {Gao, Jian and Cao, Weidong and Yang, Junyi and Zhang, Xuan},
  title     = {{AnalogGenie}: A Generative Engine for Automatic Discovery of Analog Circuit Topologies},
  booktitle = {The Thirteenth International Conference on Learning Representations},
  year      = {2025}
}

@misc{pyspice,
  author       = {Salvaire, Fabrice},
  title        = {{PySpice}: Simulate Electronic Circuit using Python and the {Ngspice}/{Xyce} Simulators},
  year         = {2021},
  version      = {1.5},
  howpublished = {Software},
  url          = {https://pyspice.fabrice-salvaire.fr/},
  note         = {Accessed 2026-07-28}
}

@article{cordella2004vf2,
  author  = {Cordella, Luigi P. and Foggia, Pasquale and Sansone, Carlo and Vento, Mario},
  title   = {A (Sub)Graph Isomorphism Algorithm for Matching Large Graphs},
  journal = {IEEE Transactions on Pattern Analysis and Machine Intelligence},
  year    = {2004},
  volume  = {26},
  number  = {10},
  pages   = {1367--1372},
  doi     = {10.1109/TPAMI.2004.75}
}

@article{wilson1927probable,
  author  = {Wilson, Edwin B.},
  title   = {Probable Inference, the Law of Succession, and Statistical Inference},
  journal = {Journal of the American Statistical Association},
  year    = {1927},
  volume  = {22},
  number  = {158},
  pages   = {209--212},
  doi     = {10.1080/01621459.1927.10502953}
}

@inproceedings{wei2022cot,
  author    = {Wei, Jason and Wang, Xuezhi and Schuurmans, Dale and Bosma, Maarten and Ichter, Brian and Xia, Fei and Chi, Ed H. and Le, Quoc V. and Zhou, Denny},
  title     = {Chain-of-Thought Prompting Elicits Reasoning in Large Language Models},
  booktitle = {Advances in Neural Information Processing Systems},
  year      = {2022},
  volume    = {35},
  pages     = {24824--24837}
}

@misc{illusiondiminishingreturns,
  author        = {Sinha, Akshit and Arun, Arvindh and Goel, Shashwat and Staab, Steffen and Geiping, Jonas},
  title         = {The Illusion of Diminishing Returns: Measuring Long Horizon Execution in {LLM}s},
  year          = {2025},
  eprint        = {2509.09677},
  archivePrefix = {arXiv},
  primaryClass  = {cs.AI}
}

@article{chen2021codex,
  title   = {Evaluating Large Language Models Trained on Code},
  author  = {Chen, Mark and Tworek, Jerry and Jun, Heewoo and Yuan, Qiming
             and Pinto, Henrique Ponde de Oliveira and Kaplan, Jared
             and Edwards, Harri and Burda, Yuri and Joseph, Nicholas
             and Brockman, Greg and others},
  journal = {arXiv preprint arXiv:2107.03374},
  year    = {2021}
}

@inproceedings{jimenez2024swebench,
  title     = {{SWE-bench}: Can Language Models Resolve Real-World {GitHub} Issues?},
  author    = {Jimenez, Carlos E. and Yang, John and Wettig, Alexander
               and Yao, Shunyu and Pei, Kexin and Press, Ofir
               and Narasimhan, Karthik},
  booktitle = {International Conference on Learning Representations},
  year      = {2024}
}

\clearpage
\appendix
\section{Additional Results}
\label{app:additional-results}

\begin{table}[H]
\centering
\caption{Task-level pass rates for SPICE and PySpice.}
\label{tab:spice_pyspice_results}
\scriptsize
\setlength{\tabcolsep}{3.2pt}
\renewcommand{\arraystretch}{0.92}
\begin{tabular*}{\columnwidth}{@{\extracolsep{\fill}}lcccc@{}}
      \toprule
      Task & DS-S & DS-Py & Qwen-S & Qwen-Py \\
      \midrule
      Connectivity edit & 0.80 & \textbf{1.00} & \textbf{0.60} & 0.57 \\
      Device add & 0.50 & \textbf{0.80} & \textbf{0.50} & \textbf{0.50} \\
      Device remove & \textbf{1.00} & \textbf{1.00} & \textbf{1.00} & 0.97 \\
      Device replace & 0.70 & \textbf{0.80} & \textbf{0.80} & 0.73 \\
      Parameter edit & 0.93 & \textbf{0.97} & \textbf{0.97} & 0.93 \\
      Rename propagation & 0.90 & \textbf{0.93} & \textbf{0.93} & 0.77 \\
      Compound 3-step & 0.13 & \textbf{0.50} & 0.20 & \textbf{0.23} \\
      Compound 6-step & 0.13 & \textbf{0.23} & 0.07 & \textbf{0.20} \\
      Compound 9-step & 0.03 & \textbf{0.23} & \textbf{0.07} & 0.03 \\
      Compound 12-step & 0.00 & \textbf{0.10} & 0.00 & \textbf{0.03} \\
      Compound 15-step & 0.00 & \textbf{0.07} & \textbf{0.00} & \textbf{0.00} \\
      Subckt inline-expand & 0.37 & \textbf{0.47} & 0.13 & \textbf{0.37} \\
      Subckt port-swap & 0.60 & \textbf{0.73} & 0.27 & \textbf{0.33} \\
      Subckt compound 3-step & \textbf{0.33} & 0.10 & \textbf{0.37} & 0.10 \\
      Subckt compound 6-step & \textbf{0.07} & 0.03 & 0.00 & \textbf{0.03} \\
      Subckt compound 9-step & \textbf{0.03} & 0.00 & \textbf{0.03} & 0.00 \\
      \midrule
      \textit{Edit subtotal} & 0.41 & \textbf{0.50} & \textbf{0.37} & 0.36 \\
      \midrule
      Equivalence judgment & 0.50 & \textbf{0.53} & 0.50 & \textbf{0.60} \\
      Device parameter & \textbf{1.00} & \textbf{1.00} & \textbf{1.00} & 0.90 \\
      Semantic terminal conn. & 0.80 & \textbf{0.83} & 0.23 & \textbf{0.43} \\
      Ordered terminal conn. & \textbf{1.00} & \textbf{1.00} & \textbf{1.00} & 0.97 \\
      Node incidence & 0.53 & \textbf{0.60} & \textbf{0.23} & \textbf{0.23} \\
      Subckt port list & \textbf{1.00} & \textbf{1.00} & \textbf{1.00} & \textbf{1.00} \\
      Instance port map & \textbf{0.90} & \textbf{0.90} & 0.80 & \textbf{0.93} \\
      Terminal neighbor inc. & 0.07 & \textbf{0.13} & \textbf{0.03} & 0.00 \\
      \midrule
      \textit{Recognition subtotal} & 0.76 & \textbf{0.78} & 0.61 & \textbf{0.64} \\
      \midrule
      \textit{Overall} & 0.52 & \textbf{0.58} & 0.45 & \textbf{0.45} \\
      \bottomrule
    \end{tabular*}

\raggedright\tiny
DS-S: DeepSeek-V4 SPICE; DS-Py: DeepSeek-V4 PySpice;
Qwen-S: Qwen-3.6 SPICE; Qwen-Py: Qwen-3.6 PySpice.
\end{table}

\begin{table}[H]
\centering
\caption{Task-level pass rates under different prompting modes.}
\label{tab:prompting_modes}
\scriptsize
\setlength{\tabcolsep}{1.8pt}
\renewcommand{\arraystretch}{0.92}
\begin{tabular*}{\columnwidth}{@{\extracolsep{\fill}}lcccccc@{}}
      \toprule
      \multirow{2}{*}{Task}
      & \multicolumn{3}{c}{DeepSeek-V4}
      & \multicolumn{3}{c}{Qwen-3.6} \\
      \cmidrule(lr){2-4} \cmidrule(lr){5-7}
      & Dir. & Think & CoT & Dir. & Think & CoT \\
      \midrule
      Connectivity edit & 0.80 & \textbf{1.00} & \textbf{1.00} & 0.60 & \textbf{0.93} & 0.87 \\
      Device add & 0.50 & 0.47 & \textbf{0.57} & 0.50 & \textbf{0.83} & 0.67 \\
      Device remove & \textbf{1.00} & \textbf{1.00} & \textbf{1.00} & \textbf{1.00} & \textbf{1.00} & \textbf{1.00} \\
      Device replace & 0.70 & \textbf{0.90} & 0.83 & 0.80 & \textbf{0.93} & \textbf{0.93} \\
      Parameter edit & 0.93 & 0.97 & \textbf{1.00} & \textbf{0.97} & \textbf{0.97} & \textbf{0.97} \\
      Rename propagation & 0.90 & \textbf{1.00} & \textbf{1.00} & 0.93 & 0.97 & \textbf{1.00} \\
      Compound 3-step & 0.13 & \textbf{0.57} & 0.50 & 0.20 & \textbf{0.73} & \textbf{0.73} \\
      Compound 6-step & 0.13 & 0.50 & \textbf{0.53} & 0.07 & \textbf{0.67} & 0.63 \\
      Compound 9-step & 0.03 & 0.47 & \textbf{0.57} & 0.07 & \textbf{0.73} & 0.70 \\
      Compound 12-step & 0.00 & \textbf{0.37} & 0.27 & 0.00 & \textbf{0.63} & 0.33 \\
      Compound 15-step & 0.00 & \textbf{0.18} & 0.14 & 0.00 & \textbf{0.54} & 0.32 \\
      Subckt inline-expand & 0.37 & 0.97 & \textbf{1.00} & 0.13 & \textbf{0.83} & 0.77 \\
      Subckt port-swap & 0.60 & \textbf{1.00} & \textbf{1.00} & 0.27 & \textbf{0.93} & \textbf{0.93} \\
      Subckt compound 3-step & 0.33 & \textbf{0.73} & 0.63 & 0.37 & \textbf{0.90} & 0.63 \\
      Subckt compound 6-step & 0.07 & \textbf{0.60} & 0.53 & 0.00 & \textbf{0.77} & 0.70 \\
      Subckt compound 9-step & 0.03 & \textbf{0.63} & 0.40 & 0.03 & \textbf{0.63} & 0.40 \\
      \midrule
      \textit{Edit subtotal} & 0.41 & \textbf{0.71} & 0.69 & 0.37 & \textbf{0.81} & 0.73 \\
      \midrule
      Equivalence judgment & 0.50 & \textbf{0.97} & 0.93 & 0.50 & \textbf{0.73} & 0.70 \\
      Device parameter & \textbf{1.00} & \textbf{1.00} & \textbf{1.00} & \textbf{1.00} & \textbf{1.00} & \textbf{1.00} \\
      Semantic terminal conn. & 0.80 & \textbf{1.00} & 0.93 & 0.23 & \textbf{0.90} & 0.87 \\
      Ordered terminal conn. & \textbf{1.00} & \textbf{1.00} & \textbf{1.00} & \textbf{1.00} & \textbf{1.00} & \textbf{1.00} \\
      Node incidence & 0.53 & \textbf{1.00} & 0.73 & 0.23 & \textbf{0.93} & 0.83 \\
      Subckt port list & \textbf{1.00} & \textbf{1.00} & \textbf{1.00} & \textbf{1.00} & \textbf{1.00} & 0.97 \\
      Instance port map & 0.90 & \textbf{1.00} & 0.97 & 0.80 & \textbf{1.00} & \textbf{1.00} \\
      Terminal neighbor inc. & 0.07 & \textbf{1.00} & 0.47 & 0.03 & \textbf{0.83} & 0.73 \\
      \midrule
      \textit{Recognition subtotal} & 0.76 & \textbf{1.00} & 0.87 & 0.61 & \textbf{0.95} & 0.91 \\
      \midrule
      \textit{Overall} & 0.52 & \textbf{0.81} & 0.75 & 0.45 & \textbf{0.85} & 0.78 \\
      \bottomrule
    \end{tabular*}

\raggedright\tiny
Dir.: direct prompting; Think: think-mode prompting; CoT: chain-of-thought prompting.
\end{table}

\newpage

\begin{figure}[H]
\centering
\begin{minipage}[t]{0.48\columnwidth}
  \centering
  \includegraphics[width=\linewidth]{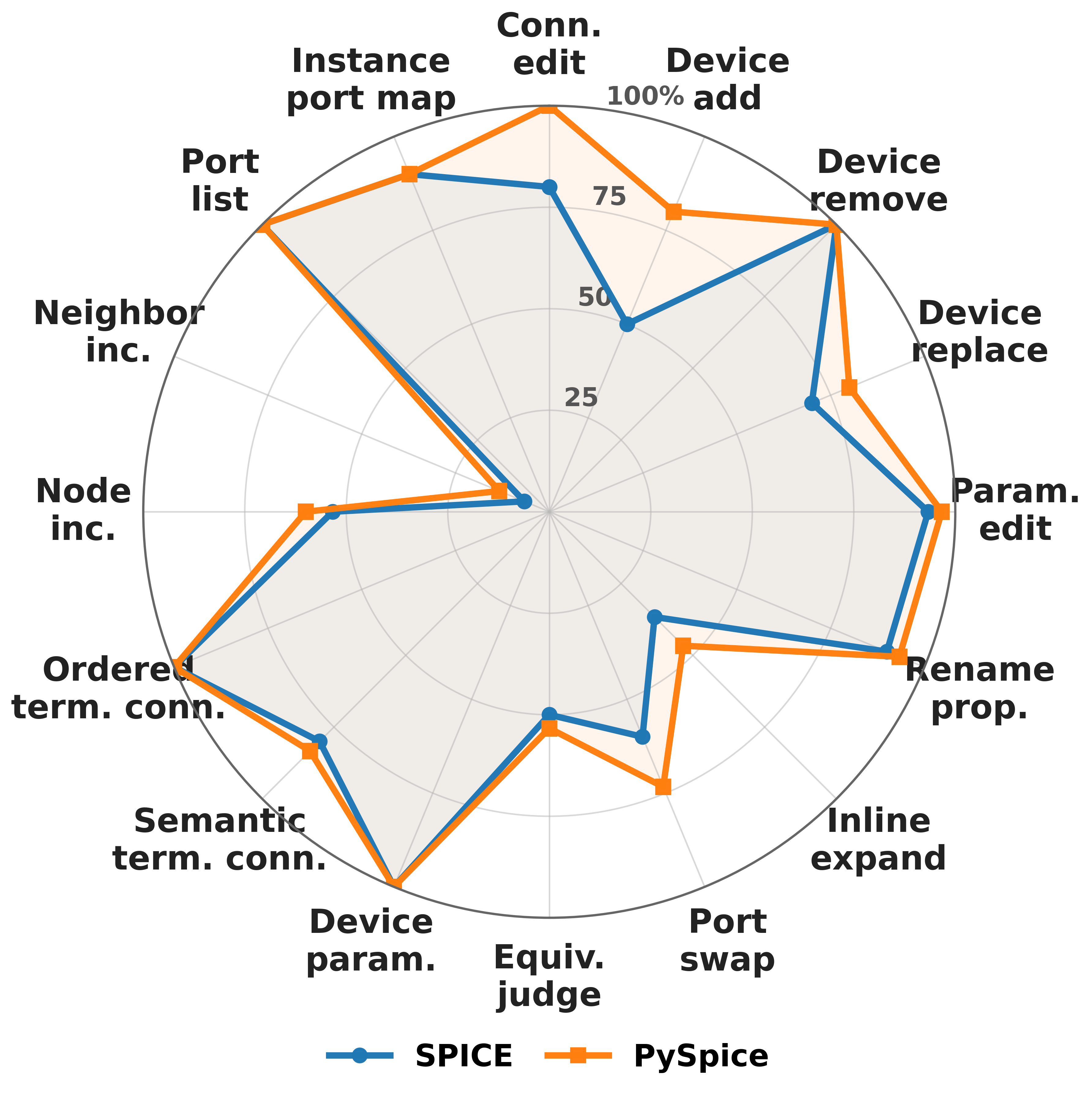}
  \centerline{\scriptsize (a) DeepSeek}
\end{minipage}\hfill
\begin{minipage}[t]{0.48\columnwidth}
  \centering
  \includegraphics[width=\linewidth]{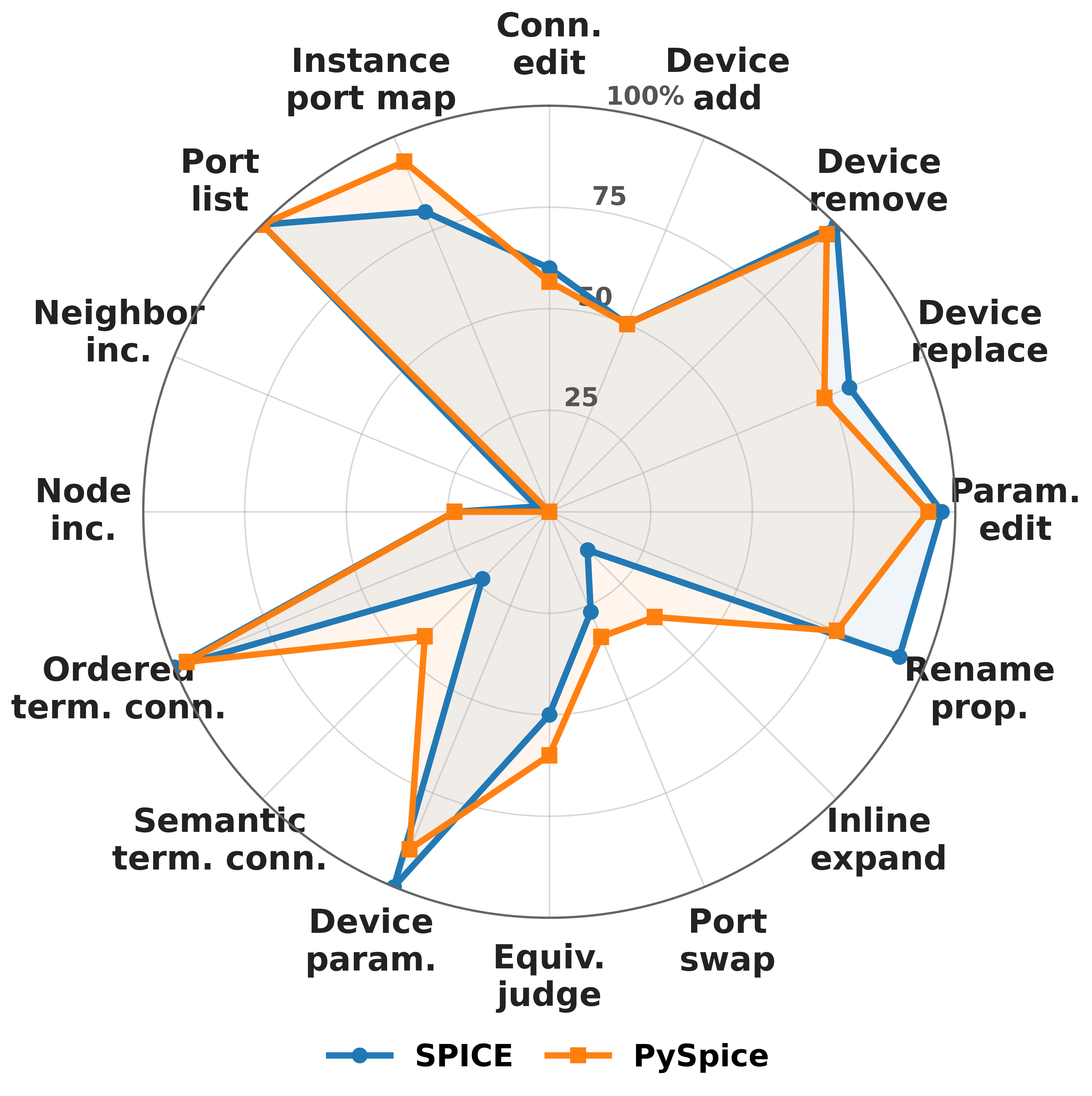}
  \centerline{\scriptsize (b) Qwen}
\end{minipage}

\begin{minipage}[t]{0.48\columnwidth}
  \centering
  \includegraphics[width=\linewidth]{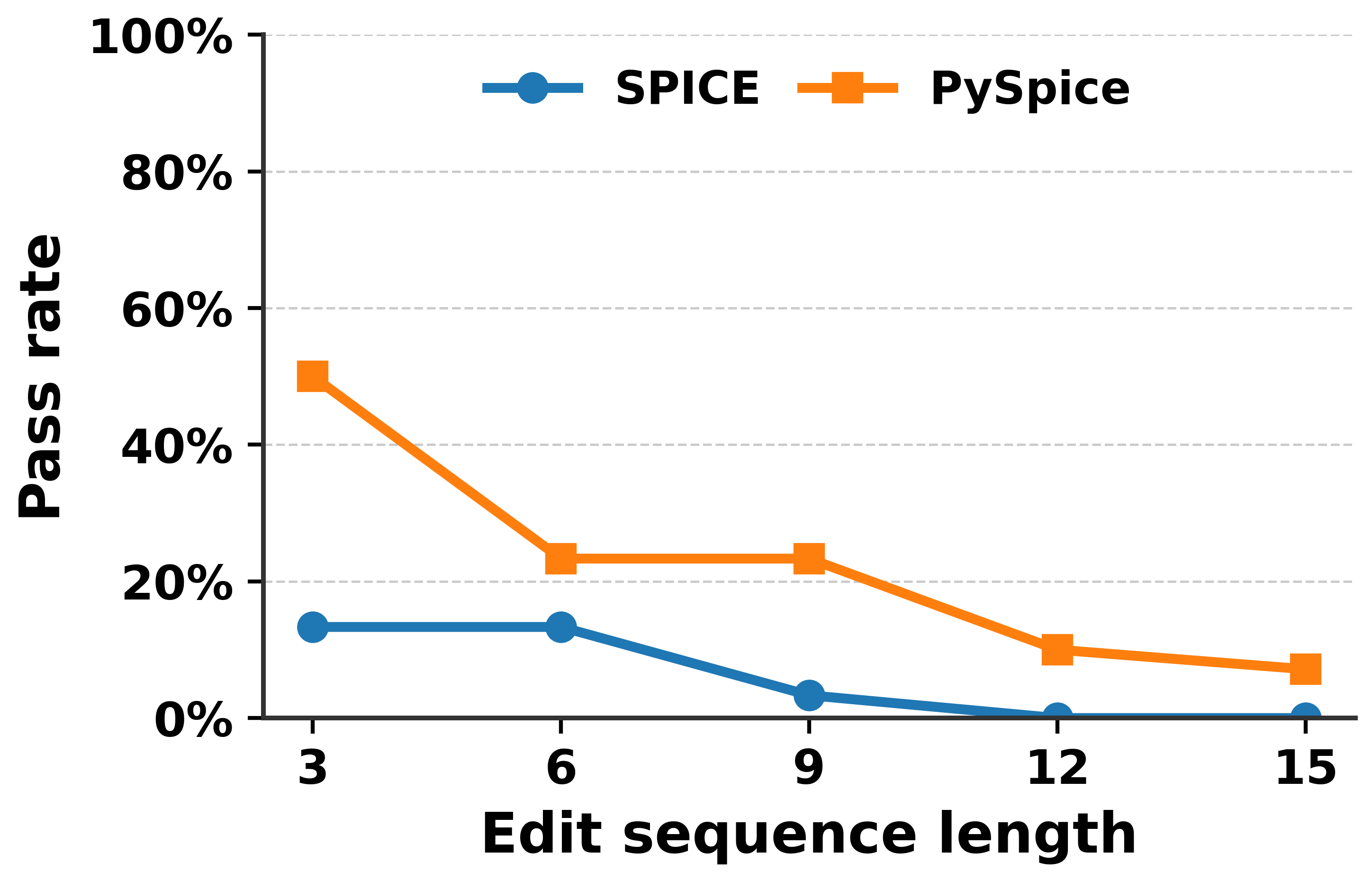}
  \centerline{\scriptsize (c) DeepSeek, Compound}
\end{minipage}\hfill
\begin{minipage}[t]{0.48\columnwidth}
  \centering
  \includegraphics[width=\linewidth]{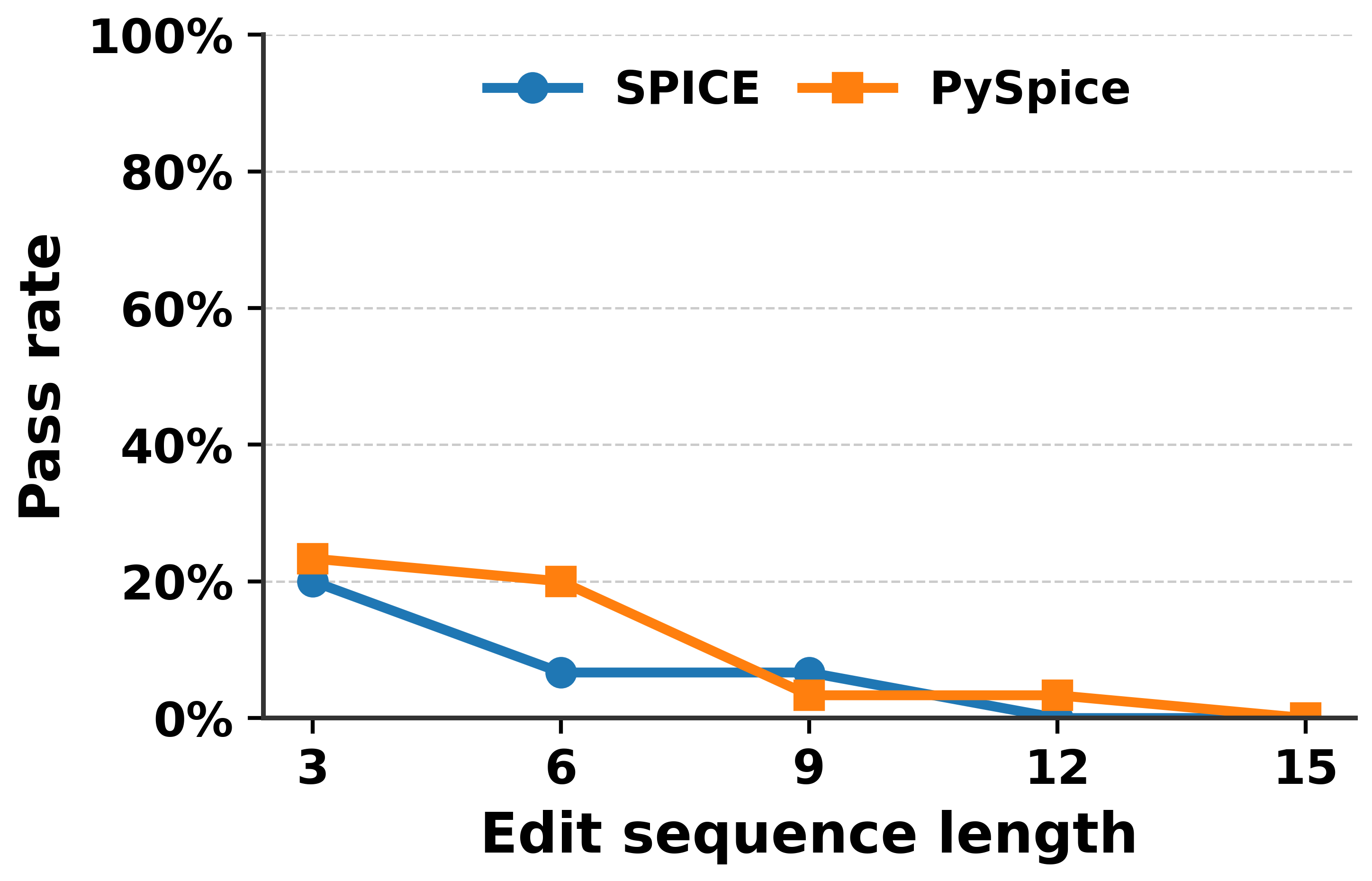}
  \centerline{\scriptsize (d) Qwen, Compound}
\end{minipage}
\caption{Task-level performance under SPICE and PySpice representations.}
\Description{Radar plots compare task-level pass rates for SPICE and PySpice representations for DeepSeek and Qwen. The lower plots show pass rates as compound-edit sequence length increases.}
\label{fig:compound_dtc_ablation}
\end{figure}

\begin{figure}[H]
\centering
\begin{minipage}[t]{0.48\columnwidth}
  \centering
  \includegraphics[width=\linewidth]{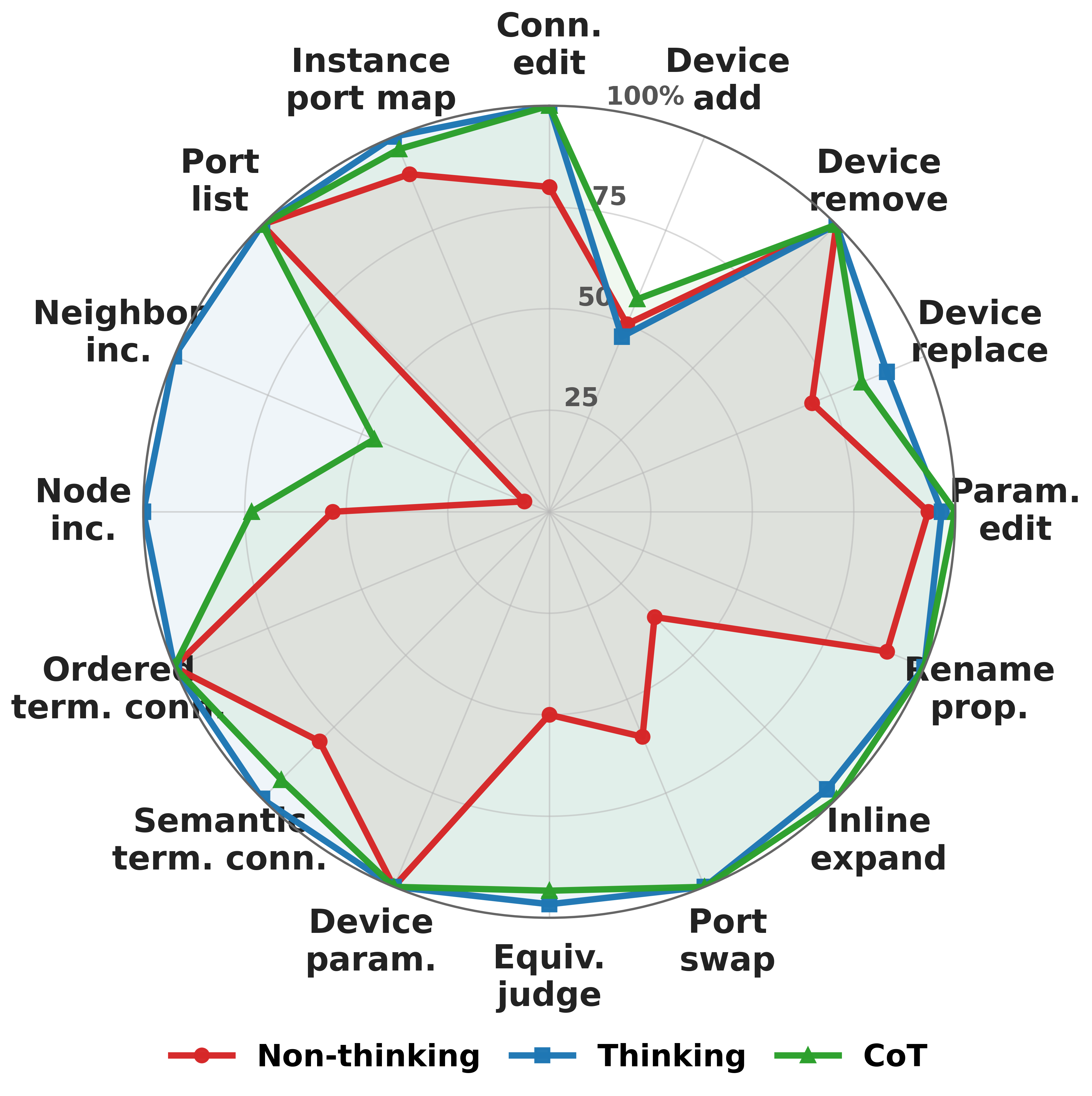}
  \centerline{\scriptsize (a) DeepSeek}
\end{minipage}\hfill
\begin{minipage}[t]{0.48\columnwidth}
  \centering
  \includegraphics[width=\linewidth]{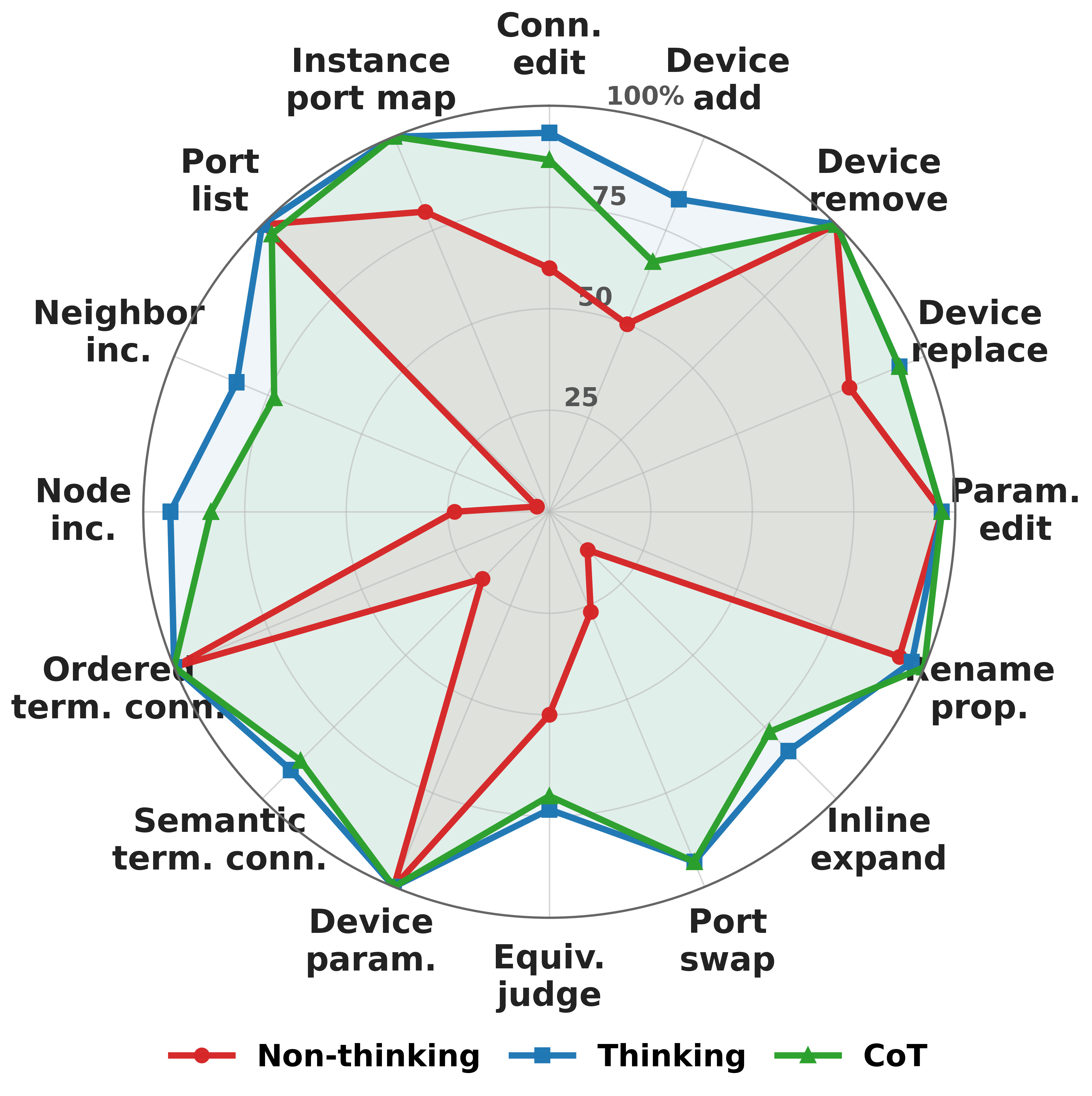}
  \centerline{\scriptsize (b) Qwen}
\end{minipage}

\begin{minipage}[t]{0.48\columnwidth}
  \centering
  \includegraphics[width=\linewidth]{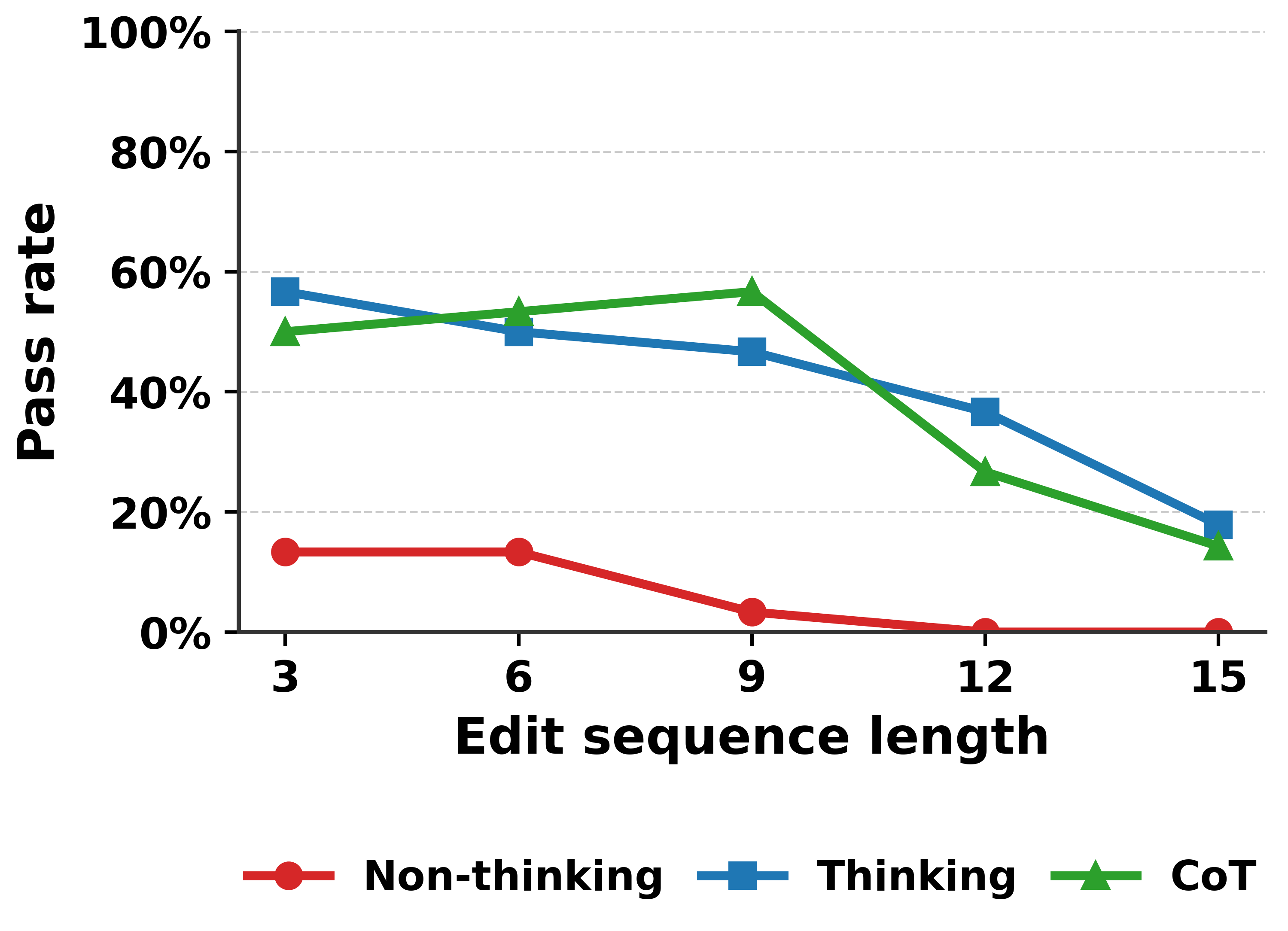}
  \centerline{\scriptsize (c) DeepSeek, Compound}
\end{minipage}\hfill
\begin{minipage}[t]{0.48\columnwidth}
  \centering
  \includegraphics[width=\linewidth]{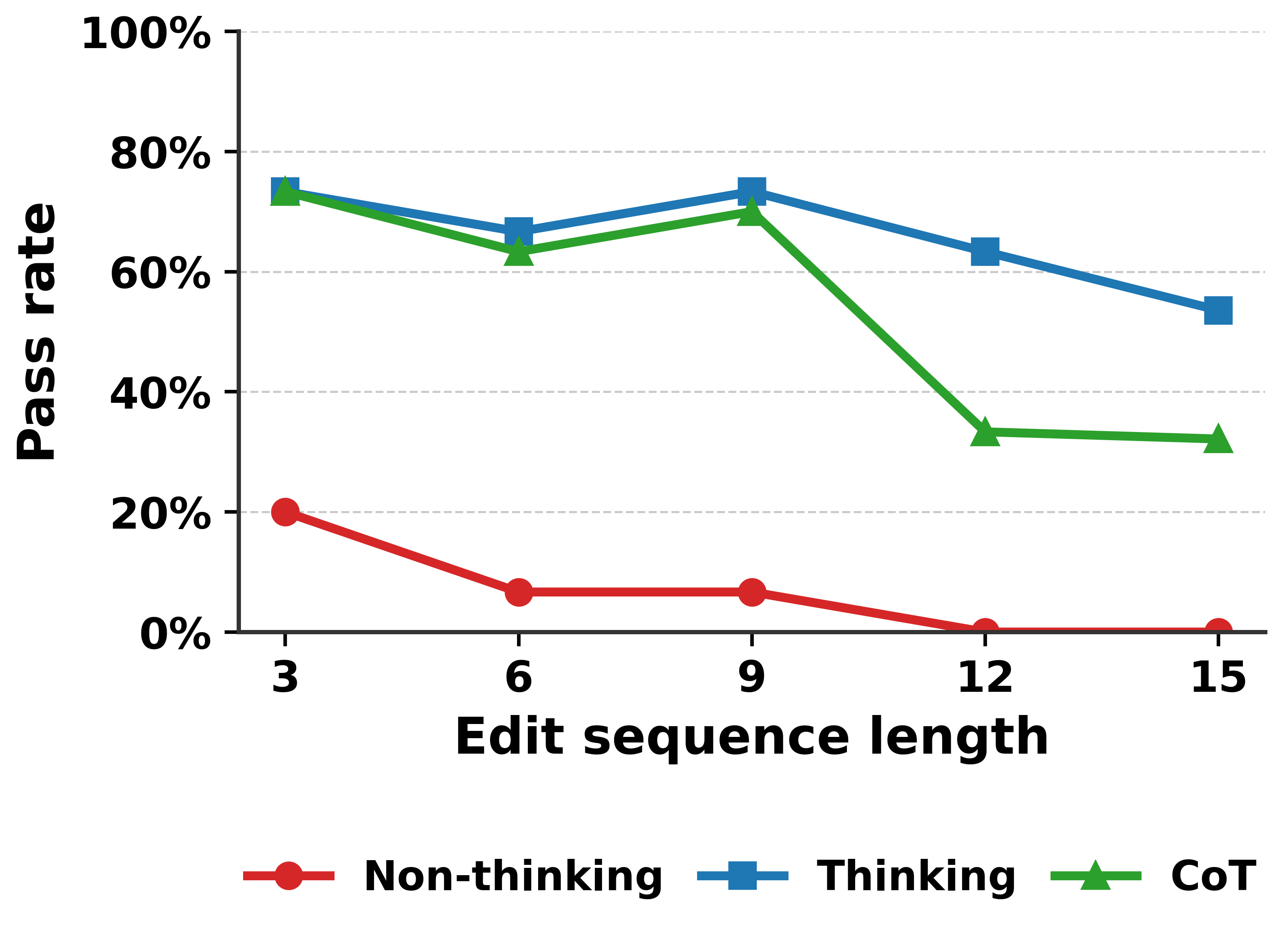}
  \centerline{\scriptsize (d) Qwen, Compound}
\end{minipage}
\caption{Task-level performance under non-thinking, thinking, and CoT prompting.}
\Description{Radar plots compare task-level pass rates under non-thinking, thinking, and chain-of-thought prompting for DeepSeek and Qwen. The lower plots show performance as compound-edit sequence length increases.}
\label{fig:radar_dtc_ablation}
\end{figure}

\end{document}